\documentclass[aps,prl,superscriptaddress,amsmath,amssymb,a4paper,twocolumn,longbibliography]{revtex4-1}

\usepackage[utf8]{inputenc}
\usepackage{amsmath}
\usepackage{amssymb,amsfonts,latexsym}
\usepackage{bm}
\usepackage[mathcal]{euscript}
\usepackage{graphicx}
\usepackage{epsfig}
\usepackage{color}
\usepackage{pifont,wasysym,marvosym}
\usepackage{textcomp}
\usepackage{comment}
\usepackage{epstopdf}
\usepackage{hyperref}
\setcitestyle{super}

\begin{document}
\title{Multi-step self-guided pathways for shape-changing metamaterials}

\author{Corentin Coulais}
\affiliation{AMOLF, Science Park 104, 1098 XG Amsterdam, The Netherlands}
\affiliation{Huygens-Kamerlingh Onnes Lab, Universiteit Leiden, PObox 9504, 2300 RA Leiden, The Netherlands}
\affiliation{Institute of Physics, Universiteit van Amsterdam, Science Park 904, 1098 XH Amsterdam, The Netherlands}
\author{Alberico Sabbadini}
\affiliation{Huygens-Kamerlingh Onnes Lab, Universiteit Leiden, PObox 9504, 2300 RA Leiden, The Netherlands}
\author{Fr\'e Vink}
\affiliation{Huygens-Kamerlingh Onnes Lab, Universiteit Leiden, PObox 9504, 2300 RA Leiden, The Netherlands}
\author{Martin van Hecke}
\affiliation{AMOLF, Science Park 104, 1098 XG Amsterdam, The Netherlands}
\affiliation{Huygens-Kamerlingh Onnes Lab, Universiteit Leiden, PObox 9504, 2300 RA Leiden, The Netherlands}

\maketitle

\textbf{Multi-step pathways, constituted of a
sequence of reconfigurations, are central to a wide variety of natural and man-made systems.
Such pathways autonomously execute in self-guided processes such as protein folding\cite{Proteinfolding} and self-assembly\cite{DNA_Selfassembly,DNA_Selfassembly_smiley,Colloidal_Selfassembly_kagome,Zeravcic_selfassemblyI},
but require external control in macroscopic mechanical systems, provided by{, e.g.,}
actuators in robotics \cite{Lipson_robotsI,Whitesides_crawlingrobot,Origami_crawlingrobot,Overvelde_origami2} or manual folding in origami\cite{Waitukaitis_origami,Origami_crawlingrobot,Murugan_SelffoldingI,Murugan_SelffoldingII}.
Here we introduce shape-changing mechanical metamaterials,
that exhibit self-guided multi-step pathways in response to global uniform compression.
Their design combines strongly nonlinear mechanical elements with a multimodal architecture that allows for a sequence of topological reconfigurations, { i.e., modifications of the} topology caused by the formation of internal self-contacts. We realized such metamaterials by digital manufacturing, and show that the pathway and final configuration can be controlled by rational design of the nonlinear mechanical elements. We furthermore
demonstrate that self-contacts suppress pathway errors.
Finally, we demonstrate how hierarchical architectures allow to extend the number of distinct reconfiguration steps. Our work establishes general principles for designing mechanical pathways, {opening} new avenues for self-folding media\cite{Murugan_SelffoldingI,Murugan_SelffoldingII}, pluripotent materials\cite{Cho_fractal,Overvelde_origami2}, and pliable devices\cite{Rogers_Science2015} in, e.g., stretchable electronics and soft robotics\cite{Review_selffolding}.}

\begin{figure*}[t!]
\begin{center}
\includegraphics[width=.99\linewidth,clip,trim=0cm 0cm 0cm 0cm]{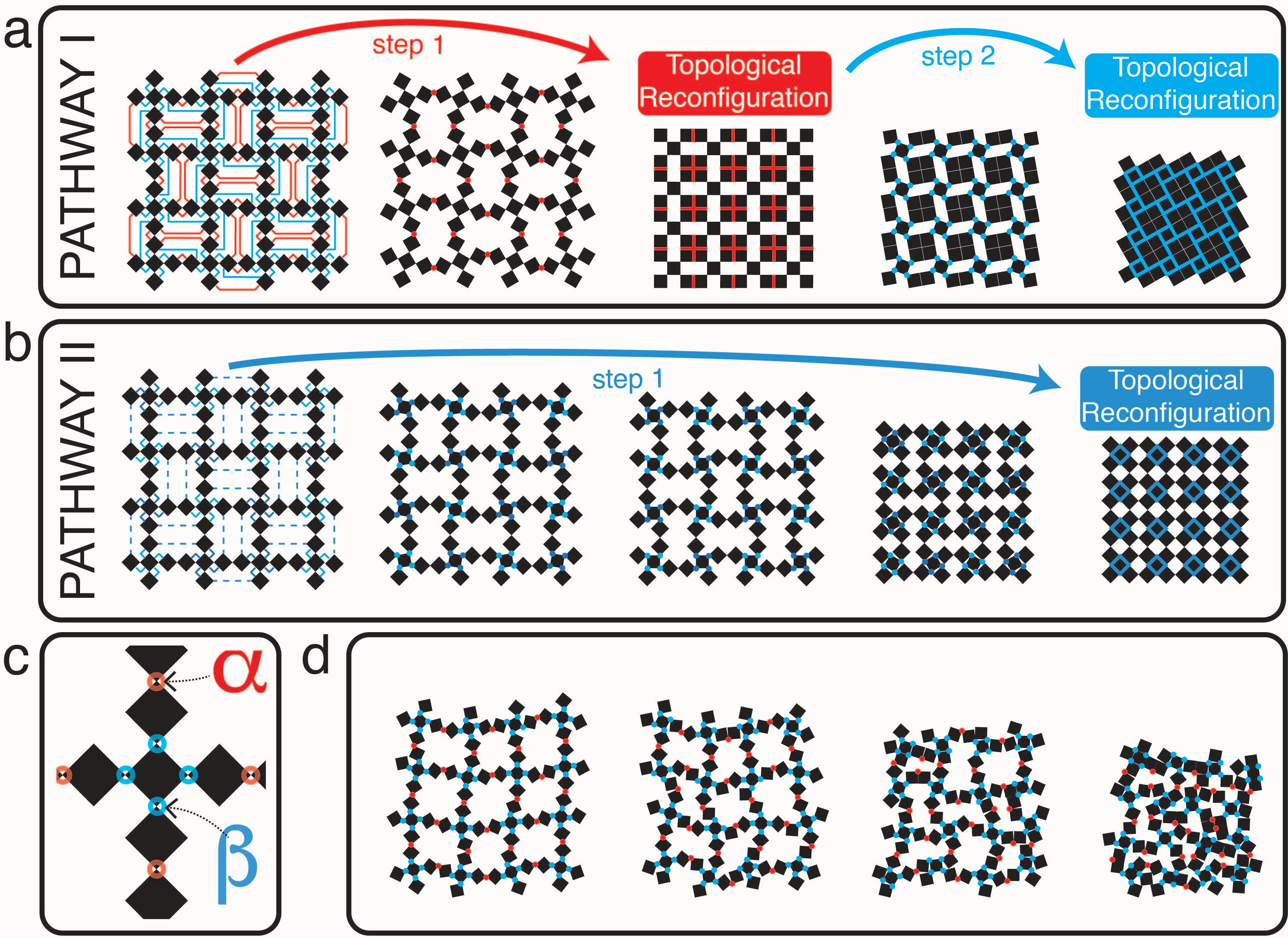}
\end{center}
\label{fig1}
\caption{{\bf Self-guided pathways in shape-changing metamaterials.} (a) Two-step pathway I. In the left panel, the red and blue lines represent the network of self-contacts formed during compression. In the first step, the $\alpha$-links (indicated by red dots) fold and the squares connected by red lines form self-contacts {(indicated in red)}, leading to the first topological reconfiguration; in the second step, the $\beta$-links (blue dots) fold and the squares connected by blue lines form additional self-contacts {(indicated in blue)}, leading to the second topological reconfiguration. (b) Pathway II is a single-step pathway, where the $\alpha$ links remain straight and a topological reconfiguration (blue lines) occurs after the $\beta$-links are maximally folded. (c) Zoom-in of the unit cell; red and blue circles indicate the $\alpha$ and $\beta$ links and the angles $\alpha$ and $\beta$ are defined as the deviation from the undeformed state. (d) Example of a disordered pathway.}
\end{figure*}

Mechanical metamaterials are structured forms of matter that can be designed to exhibit a wide range of anomalous properties\cite{Bertoldi_NatRevMat2017}, including negative responses\cite{Bertoldi_poisson}, topological polarization\cite{Huber_reviewtopological}, non-reciprocity\cite{Coulais_nonreci} and shape morphing \cite{Coulais_metacube,Overvelde_origami2}. These striking deviations from ordinary elasticity ultimately emerge from soft deformation modes, encoded in their internal architecture \cite{Bertoldi_NatRevMat2017,Huber_reviewtopological}. Most metamaterials feature a single soft mode, which limits their deformations to single-step  pathways \cite{Grima_squares,Mullin_holey,Coulais_metacube,Coulais_Natphys2017}. Multi-step deformations, as well as other advanced functionalities, require designs featuring a high dimensional deformation space spanned by multiple soft modes \cite{Cho_fractal,Grima_Hierarchical,ShuYang_Hierarchical,Bertoldi_Hierarchical,Overvelde_origami1,Overvelde_origami2}. However, when actuated, competition between those soft modes leads to frustration \cite{Murugan_SelffoldingI,Murugan_SelffoldingII} and spatial decay of functionality \cite{Coulais_Natphys2017}. Hence, multi-modal mechanical metamaterials are sofar either actuated without {explicit control over} their reconfiguration pathways\cite{Cho_fractal,ShuYang_Hierarchical}, or require the use of  multiple actuators, one for each degree of freedom\cite{Overvelde_origami1,Overvelde_origami2}. In contrast, processes  such as self-assembly\cite{Zeravcic_selfassemblyI} and protein folding\cite{Proteinfolding} exhibit robust multi-step pathways without the need for such external control. Here we raise the question whether one can design and create %multi-modal
mechanical metamaterials that translate global forcing into a self-guided multi-step pathway of reconfigurations.

To address this challenge, we consider two-dimensional metamaterials consisting of diluted lattices of freely hinged squares. We design a unit cell that consist of a cross-shaped pattern of five squares, leading to a multi-modal metamaterial of $n\times n \times 5$ squares, {which
allows for the formation of self-contacts between initially separated elements and}
features $3n^2 + 4n -3$ zero energy modes  (Fig.~1 and Methods).
%This structural design allows for the formation of self-contacts between initially separated elements when sufficiently compressed.
Crucially, this structural design admits a two-step deformation path{way}, characterized by the deformation angles $\alpha$ and $\beta$ (Fig.~1a,c). In the first step of this {pathway I}, the $\alpha$-links should fold while the $\beta$-links should remain fixed until self-contacts, which change the topology of the material,  are formed. This topological reconfiguration spawns a daughter structure at finite compression. In the second step, the $\alpha$-links should remain fixed and
the $\beta$-links should fold  until the system is fully compacted (Fig.~1a).
This is but one possible deformation path, however, and our structure admits a myriad of other pathways, some highly structured (Fig.~1b and Methods), most disordered (Fig.~1d).

To obtain a specific deformation path in response to global compression, we augment this structural design with suitably tailored mechanical interactions, connecting the squares with beam-like links of varying thickness $t_{\alpha}$ and $t_{\beta}$ (Methods). This replaces the zero energy modes by soft modes, whose energetics are controlled by the links. Crucially, these beams also introduce a well defined critical compressive load $\sim t^3$ above which the non-linear buckling instability causes spontaneous bending of the corresponding link\cite{Coulais_Metabeam}. {Intuitively, pathway I requires that
$t_{\alpha} <  t_{\beta}$, whereas pathway II requires $t_{\beta} <  t_{\alpha}$ (Fig.~1)}. The mode structure in {the corresponding } asymptotic limits reveals {an important} qualitative difference: whereas the undeformed metamaterial has a single soft mode for $t_{\alpha} \ll  t_{\beta}$, it has many (precisely $n^2+6n-3$) soft modes for $t_{\beta} \ll  t_{\alpha}$, as the numerous soft $\beta$ links can deform in myriad manners in this case (Methods). This counting argument suggests that pathway I {will be} much easier to realize than pathway II: we refer to pathway I as a natural pathway.

To rationally design the links to obtain either pathway I or II, we have analysed the modes and instabilities of a representative volume element, a so-called super cell consisting of a $2\times2$ grid of unit cells, as function of $t_{\alpha}$ and $t_{\beta}$ (Methods). {This modal analysis substantiates our counting-based argument.}
{First, we
} find that the ratio $t_{\alpha}/ t_{\beta}$, rather than the individual thicknesses, plays the crucial role. {Moreover, f}or small $t_{\alpha}/ t_{\beta}$, the first buckling mode corresponds to the first step in pathway I, with other modes strongly suppressed. {Finally,} for large $t_{\alpha}/ t_{\beta} $, even though the first buckling mode coincides with the first step in pathway II, other modes remain in competition. {
Based on the details of this analysis we extract specific design} guidelines for the rational design of the link parameters (Methods).

\begin{figure*}[t!]
\begin{center}
\includegraphics[width=.99\linewidth,clip,trim=0cm 0cm 0cm 0cm]{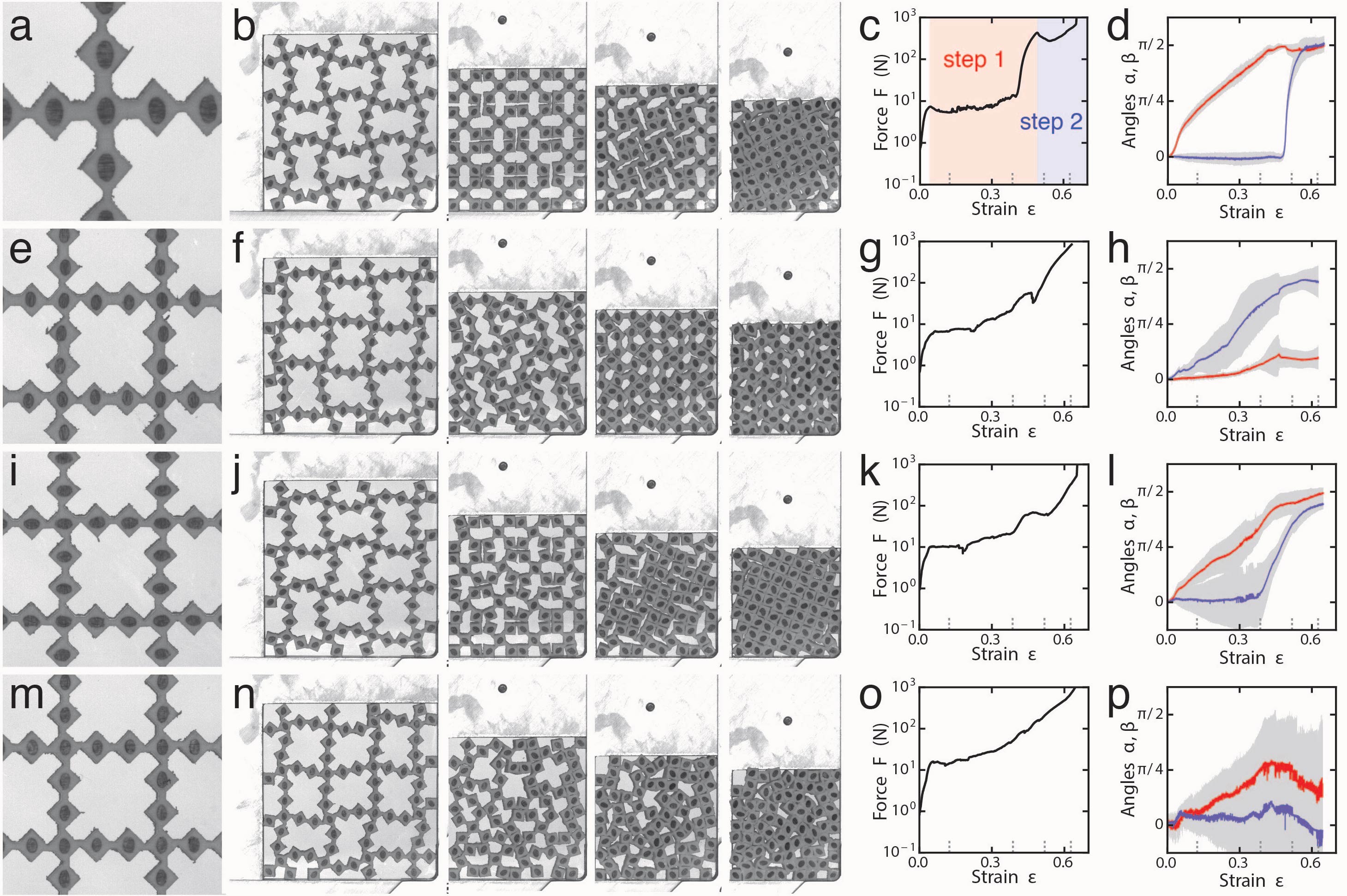}%Or v4, see which one you prefer
\end{center}
\label{fig2}
\caption{{\bf Pathway control by rational hinge design.} (a-d) Two-step pathway I; (e-h) Pathway II; (i-l) Smoothed pathway I; (m-p) Disordered pathway. (a,e,i,m) Zoom-in showing hinge designs. (b,f,j,n) A series of snapshots of the experimentally observed mechanical pathway, at engineering strains $\varepsilon= 0.12$, $0.39$ $0.52$ and $0.63$. These strains are also indicated in panels (c,d,g,h,k,l,o,p). (c,g,k,o) The compressive force vs. engineering strain $\varepsilon$; dashed ticks indicate the strains of the snapshots in panels (b,f,j,n). In {(c), we indicate the two steps of the pathway.} (d,h,l,p) Bending of the links $\alpha$, $\beta$ vs. strain $\varepsilon$ in the center region of the sample (see Methods). See also Supplementary Video 1. The solid curves indicate the mean value of $\alpha$ (red) and $\beta$ (blue), whereas the gray zones indicate one standard deviation.}
\end{figure*}

We illustrate our approach experimentally. We used a waterjet cutter to  fabricate reconfigurable metamaterials with squares of size $4.5$~mm out of  rubber sheets that are $10$ mm thick (this prevents out of plane buckling). We then actuated their deformation pathways by applying a uniform equi-biaxial strain $\varepsilon$ while tracking the compressive force, the motion of each element and the evolution of the folding angles $\alpha$ and $\beta$ (see Methods). We focus on four different samples that share the same structural design, but differ in their $\alpha$ and $\beta$ links, which are designed to obtain one of four specific pathways (Fig.~2a,e,i,m).

\begin{figure*}[t!]
\begin{center}
\includegraphics[width=\linewidth,clip,trim=0cm 0cm 0cm 0cm]{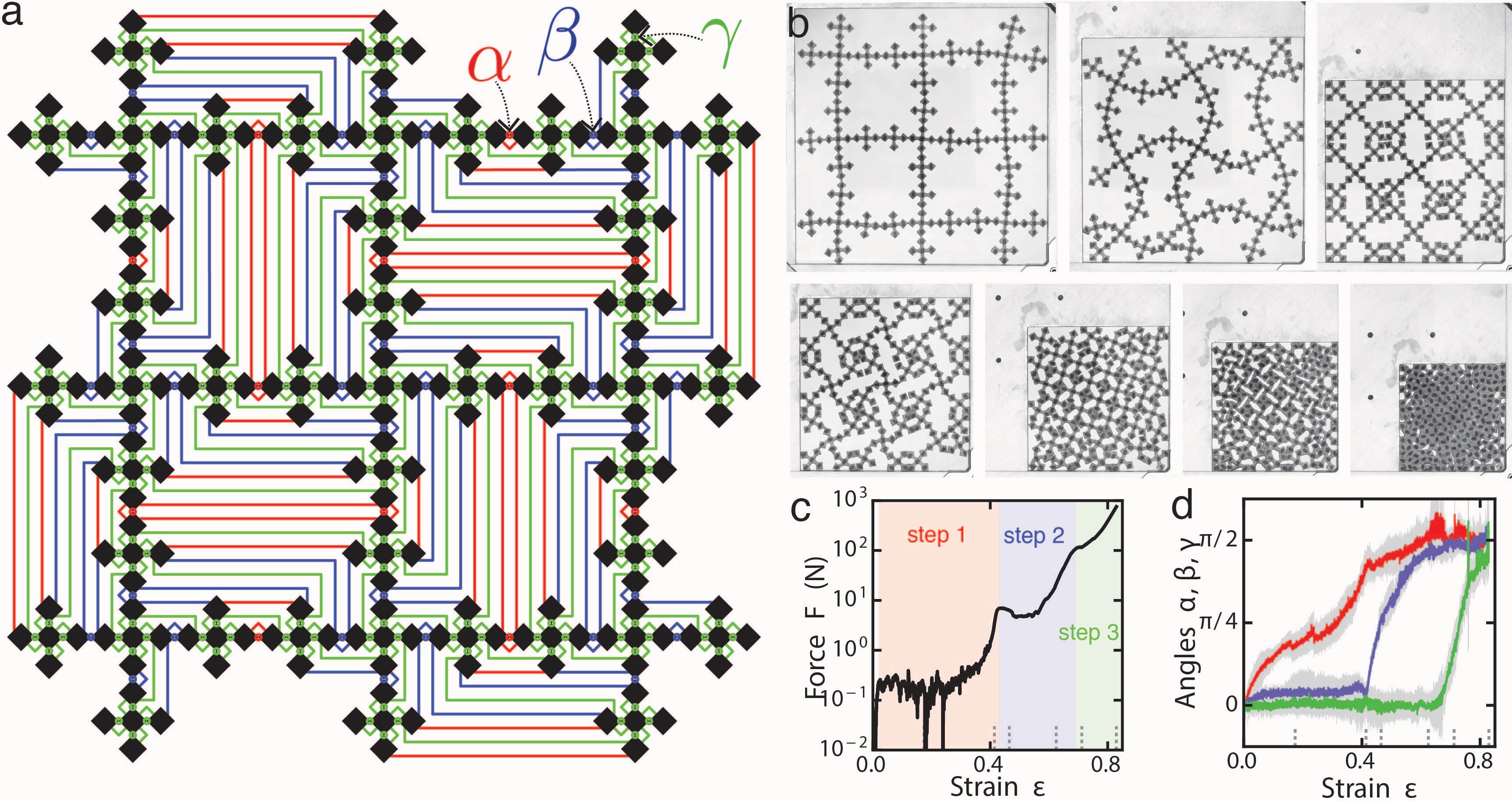}
\end{center}
\label{fig3}
\caption{{\bf Three-step mechanical pathway}.
(a) Geometry of a shape-changing hierarchical metamaterial. The red, blue and green coloured circles indicate the $\alpha$, $\beta$ and $\gamma$ links, respectively.
The red, blue and green coloured lines indicate the self-contacts between the squares (indicated in black) that form upon the subsequent folding of the $\alpha$, $\beta$ and $\gamma$ links.
(b) Snapshots of the experimentally observed three-step mechanical pathway at $\varepsilon=0$, $0.20$, $0.37$, $0.58$, $0.62$, $0.70$ and $0.85$, see also Supplementary Video 2.
%In the first zoom-in, The $\alpha$, $\beta$ and $\gamma$ links are respectively highlighted by red, blue and green circles. (b) Zoom-in of the central region of fully compressed final state (for $\varepsilon=0.85$).
(c) Compression force vs. strain. (d) Bending angles of the links $\alpha$ (red), $\beta$ (blue)  and $\gamma$ (green) vs. strain $\varepsilon$.}
\end{figure*}

We first design a metamaterial that follows the two-step pathway I. Here, following
our design guidelines (Methods) we take $t_\alpha =  1.0 $ mm and $   t_\beta = 2.0$ mm (Fig.~2a).
Under compression, the $\alpha$-links buckle at a strain of $4\%$, triggering the counter-rotation mode \cite{Mullin_holey} and the first step of the reconfiguration pathway (Fig. 2b). Further compression does not significantly increase the force \cite{Coulais_Metabeam}, preventing buckling of the stronger $\beta$ links, until self-contacts are formed at a strain of $49\%$. The ensuing topological reconfiguration gives rise to a daughter structure with only one soft mode governed by the $\beta$-links. Further compression leads to a steep rise in the compressive force which triggers the buckling of the $\beta$-links, initiating the second reconfiguration step, which continues until the structure is fully compressed (Fig. 2b). The compressive force clearly evidences the plateaus associated with buckling and the rapid increase associated with the formation of self-contacts (Fig.~2c). This multi-step pathway is fully consistent with the theoretical pathway I, as further shown by the distinct evolution of the angles $\alpha$ and $\beta$ as function of strain (Fig.~2d and Methods). Different geometries can be designed that show similar multi-step pathways under uniaxial compression (Methods). Hence, the combination of self-contacts, buckling, and topological reconfiguration enables multi-step sequences where buckling initiates bending, sufficient bending---or folding---creates self-contacts and topological reconfigurations, which in turn trigger more buckling events.

Successful execution of a structured pathway requires error-correcting mechanisms, that compensates for distortions caused by boundary frustration and manufacturing imperfections. We note that
while the scatter in $\alpha$ and $\beta$ grows significantly during each deformation step, it sharply reduces
upon reaching self-contact (Fig.~2d). This reduction results from the self-alignment of the flat interfaces of contacting  squares, and underlies the remarkable feature of our metamaterials that the fully compressed structure is highly ordered. Hence, self-contacts have a crucial error-correcting role, thus ensuring a robust execution of a pathway.

Pathway II is harder to realize experimentally.
Even for large $t_\alpha/ t_\beta =4 $ we observe that while
the deformation is qualitatively similar to pathway II, it is still quite disordered, which we attribute to the issue of competition of multiple modes (Methods).

To address this issue, we now demonstrate that symmetry breaking perturbations 
can suppress the undesired modes and {promote} pathway II.
We introduce a new design parameter $\delta_\beta$, and laterally offset the $\beta$-links over a distance $\delta_{\beta}=t_{\beta}/2$ with alternating chirality (Fig.~2e and Methods). The design guideline that we have extracted from our numerical analysis suggests that such symmetry breaking offsets are sufficient to favour pathway II already for $t_\alpha/ t_\beta \gtrsim 2$ (Methods). We thus created a second sample with such offsets, took $t_\alpha =  2.0 $ mm  $> t_\beta = 0.8$ mm, and find that under compression, the slender $\beta$-links bend, the thicker $\alpha$-links remain essentially straight, and a single-step path towards a different end state ensues (Fig.~2e-h).

We note that offsets can also be used to {modify and probe} pathway I. Fixing $\delta_\beta=t_\beta/2$ with constant chirality (Fig.~2i,m) and gradually increasing $t_{\alpha}/t_{\beta}$, we find experimentally that up to $t_{\alpha}/t_{\beta}\lesssim 1.2$, the broken symmetry causes blurring of the pathway and larger scatter in the bending angles, but {pathway I still ensues and} the fully folded state is still reached, illustrating the robustness of the natural pathway against perturbations (Fig.~2i-l; $t_\alpha =  1.2 $ mm, $t_\beta = 1.0 $ mm). For even larger $t_{\alpha}/t_{\beta}\gtrsim 1.4$, we observe a disordered pathway (Fig.~2m-p, where $t_\alpha=1.4$ mm, $t_\beta={1.0}$ mm). {These examples with offset links demonstrate that symmetry breaking provides an efficient strategy to steer the deformation pathway.}

Finally, the hierarchical nature of our structural design can be leveraged to create metamaterials with a three-step natural pathway. Our two-step design results from replacing each square in a rotating square mechanism\cite{Grima_squares} with a cross pattern of five smaller squares (Methods). To obtain a three-step sequence, we replace each square in the unit cell of the two-step design by a smaller
cross-shaped subpattern (Methods). The resulting unit cell consists of 25 small squares that form a generation-2 box fractal; we create a metamaterial out of nine of these unit cells which, in the limit of perfect hinges, has 216 zero modes (Fig.~3a and Methods). We aim to obtain a three-step pathway by sequential folding of the $\alpha$, $\beta$ and $\gamma$ links, which leads to an intricate web of self-contacts (Fig. 3a and Methods). We tailor the three classes of  links to set up such pathway by taking $t_{\alpha}=1.0$ mm $< t_{\beta}=1.2$ mm $< t_{\gamma}=2.0$ mm. Upon biaxial compression, we observe a three-step sequence of reconfigurations, leading to a fully closed, highly ordered structure (See Fig.~3b and Supplementary Video 2). The compressional force and bending angles further evidence the three distinct steps of this reconfiguration pathway (Fig.~3c-d),  demonstrating that our design strategy can naturally be used to achieve complex, multiple-step pathways that are remarkably robust.

To conclude, we have shown that hierarchical structures dressed with mechanical switches based on buckling and self-contacts feature self-guided topological reconfigurations that are robust to imperfections.
{Our designs are purely geometric and can thus be applied over a range of scales, and they are passive,
thus alleviating} the need for external control. {We expect our approach to considerably enhance the potential of,} e.g., reconfigurable materials\cite{Cho_fractal,MacEuen_GrapheneKiri,Rogers_Kiri,Overvelde_origami2}, soft robotics\cite{Origami_crawlingrobot}, origami metamaterials\cite{Waitukaitis_origami,Murugan_SelffoldingI,Murugan_SelffoldingII}, and stretchable electronics\cite{Cho_fractal,Rogers_Science2015}.

{\em Acknowledgements.} We thank J. Mesman and D. Ursem for outstanding technical support. We acknowledge Z. Zeravcic and A. Murugan for discussions.
We acknowledge funding from the Netherlands Organization for Scientific Research through grants VICI No. NWO-680-47-609 (M.v.H.) and VENI NWO-680-47-445 (C.C).

{\em Author Contribution.} All authors conceived the project. C.C. and M.v.H. developed the theoretical models and C.C. performed the numerical simulations. C.C., A.S. and F.V. designed and carried out the experiments. C.C. and M.v.H. wrote the manuscript.

\clearpage
\section{Methods}
\setcounter{figure}{0}
\renewcommand{\figurename}{Extended Data Figure}
\renewcommand\thefigure{ED\arabic{figure}}
\setcounter{table}{0}
\renewcommand{\tablename}{Extended Data Table}
\renewcommand\thetable{ED\arabic{table}}

\subsection{Hierarchical Design}
\begin{figure*}[h!]
\begin{center}
\includegraphics[width=.6\linewidth,clip,trim=0cm 0cm 0cm 0cm]{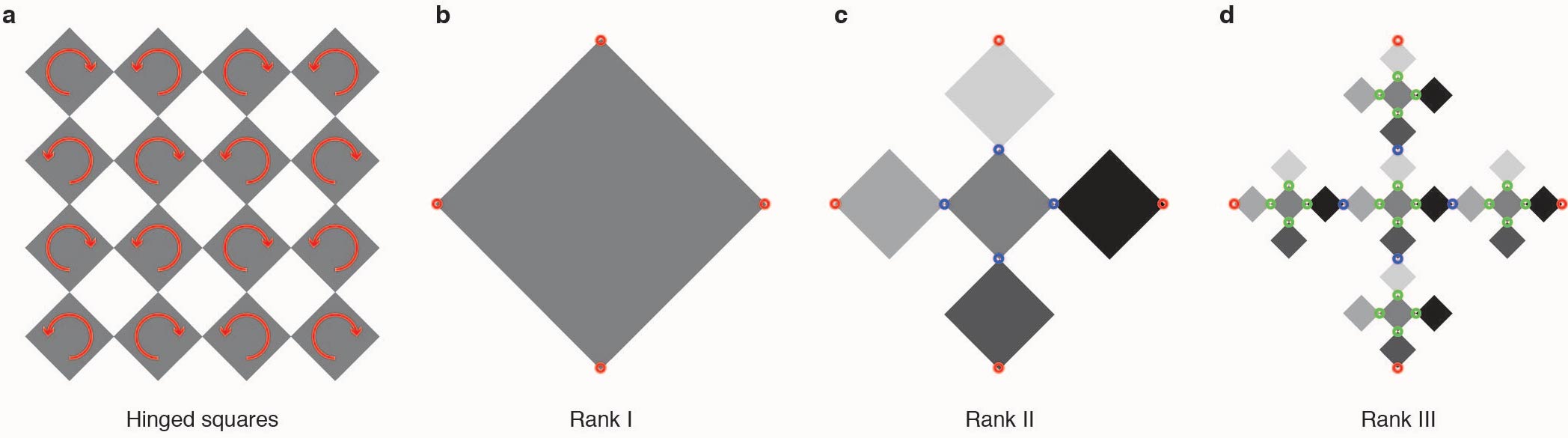}
\end{center}
\caption{{\bf Multimodal hierarchical mechanical metamaterials.} (a) $4\times 4 $ rotating square mechanism, with features a zero mode corresponding to counter-rotation of each square unit as indicated. (b-d) Hierarchical construction of a box fractal, where in each generation a square is replaced by a cross-like pattern of 5 smaller squares. Red, blue and green links correspond to $\alpha$-links that connect different units, internal $\beta$ links that occur for rank $\ge 2$ and $\gamma$ links that occur for rank $\ge 3$.}
\label{M1}
\end{figure*}

The architecture of our metamaterials is based on an $n\times n$ rotating square mechanism (Extended Data Fig.~\ref{M1}a). We see this as the first rank in a hierarchy, and for increasing rank, replace each square building block by a sub-pattern of five smaller squares; this construction is known as generating a so-called box fractal (Extended Data Fig.~\ref{M1}b-d). For an $n\times n$ pattern of units of rank $m$, the number of squares equals $n^2 \times 5^{(m-1)}$. These are connected by $2 n (n-1)$ $\alpha$-links that connect different unit cells. For $m\ge 2$ there are in addition  $4$ internal $\beta$-links per unit, and for $m\ge 3$ there are in addition $4 \times 5$ internal $\gamma$-links. In general, going from rank $m-1$ to rank $m$ yields $4 n^2 \times 5^{(m-2)}$ additional internal links. For general $m \ge 2$, this yields $2n(n-1)+4n^2 \Sigma_{i=0}^{m-2}5^i = -2n+n^2\left[1+5^{(m-1}) \right]$ connections.

\subsection{Hinged Tessellations}
\begin{figure*}[!b]
\begin{center}
\includegraphics[width=0.6\linewidth,clip,trim=0cm 0cm 0cm 0cm]{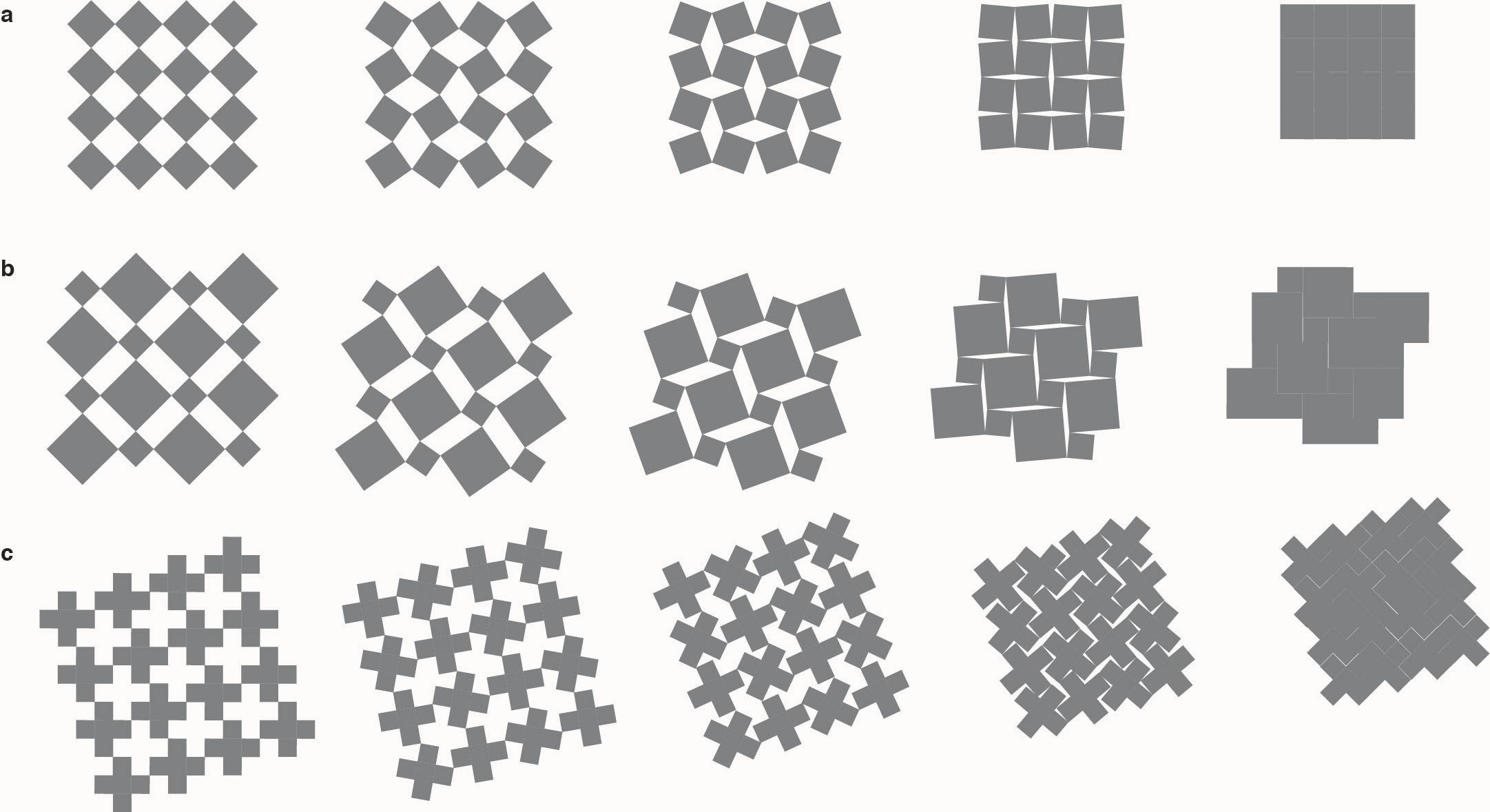}
\end{center}
\caption{{\bf Hinged tesselations.} (a) Free motion of the rotating square mechanism (b) Free motion of rotating square mechanism with unequal squares. (c) Free motion of linked crosses. }
\label{M2}
\end{figure*}

Hinged tessellations are freely hinging structures that can be folded into a fully area filing structure. The rotating square mechanism is an example of a hinged tessellation (Extended Data Fig.~\ref{M2}a). Considering pathways in our hierarchical metamaterials, we also encounter variations on the rotating square mechanism with unequal squares (Extended Data Fig.~\ref{M2}b) and  crosses (Extended Data Fig.~\ref{M2}c); both are hinged tessellations and can form area filling structures with multiple self-contacts.

\subsection{Zero Modes}
When the links between the squares are completely flexible, a simple Maxwell counting argument allows to determine the number of zero modes. We start from three degrees of freedom per square (translation and rotation), and subtract two constraints for each connection, as well as three global degrees of freedom (global translation and rotation). This yields a simple expression for $n_z$, the number of internal degrees of freedom:
\begin{equation}
n_z = n^2\left[5^{(m-1)}-2\right]+4n-3~.
\end{equation}
It can readily be verified that for $m=2$, $n_z=3n^2+4n-3$, yielding 61 internal zero modes for the $n=4$, hierarchy 2 structures that we show in Fig.~1 and 2 of the main text, and that for $m=3$, $n_z=23n^2+4n-3$, yielding 216 internal zero modes for our $n=3$, hierarchy 3 structure, shown in Fig.~3 of the main text.

To count the number of soft modes for the limit $t_{\alpha} \ll t_{\beta}$, one can consider the case that the $\beta$-links are stiff, the $\alpha$-links are floppy, resulting in a structure which is equivalent to a rotating square mechanism with one degree of freedom\cite{Grima_squares}. To count the number of soft modes for the limit $t_{\beta} \ll t_{\alpha}$, for the $m=2$ case, one can consider the case that the $\beta$-links are floppy and the $\alpha$-links are stiff; these links now impose $3$ instead of 2 constraints, so that an additional $2n(n-1)$ constraints must be subtracted from Eq. (1), yielding $n^2+6n-3$ floppy modes.

\subsection{Rational Design of Linkages}

In the metamaterial, we dress $\alpha$-links and $\beta$-links with elastic beams of thicknesses $t_{\alpha}$ and $t_{\beta}$. Such link thicknesses control the  critical buckling strains and buckling modes, with the relative thickness, $t_{\alpha}/t_{\beta}$, playing a crucial role. Our aim is to design these link parameters such that for each step of the pathway, the required mode dominates the deformations; for the specific pathways I and II, this requires to tune $t_{\alpha}$ and $t_{\beta}$ to obtain the required first buckling mode that initiates the deformation sequence under compression.

To gain insight in the role of the link thicknesses,  we take the following general design approach. First, we calculate the spectrum of soft modes at zero strain as function of $t_{\alpha}$ and $t_{\beta}$, and decompose the spatial structure of the lowest energy mode onto a basis of ``motions'', which correspond to desired (and undesired) pathway steps. Although a pathway consists of highly nonlinear deformations, we will see that such a linear analysis already provides important insights on the role of the link parameters. Second, we perform a nonlinear analysis to determine the critical strain and spatial structure of the first buckling mode, as function of the link parameters. {Altogether}, this analysis leads to rational design of the links to generate a desired deformation pathway.

{\em Motions:}
To focus on the essential physics, we define a system consisting of four unit cells, organized in a $2\times 2$ super-cell placed in a uniformly compressed square box with periodic boundary conditions. In a general $n\times n$ system, periodic boundaries tie the nodes at the left (bottom) boundary to the nodes at the right (top) boundary, leading to $4n$ constraints. To remain compatible with the experimental boundary conditions, we require that the super-cell fits in a square box: the $n$ edges at each side are required to be aligned, leading to an additional $2(n-1)$ constraints, and the horizontal and vertical dimensions are required to be equal, leading to a single additional constraint. Hence, in the limit of flexible hinges, the number of zero modes with square periodic boundary conditions becomes:
\begin{equation}
n_z^{\textrm{periodic}} = n^2\left[5^{(m-1)}-2\right]-2n-2~.
\end{equation}

\begin{figure*}[h!]
\begin{center}
\includegraphics[width=\linewidth,clip,trim=0cm 0cm 0cm 0cm]{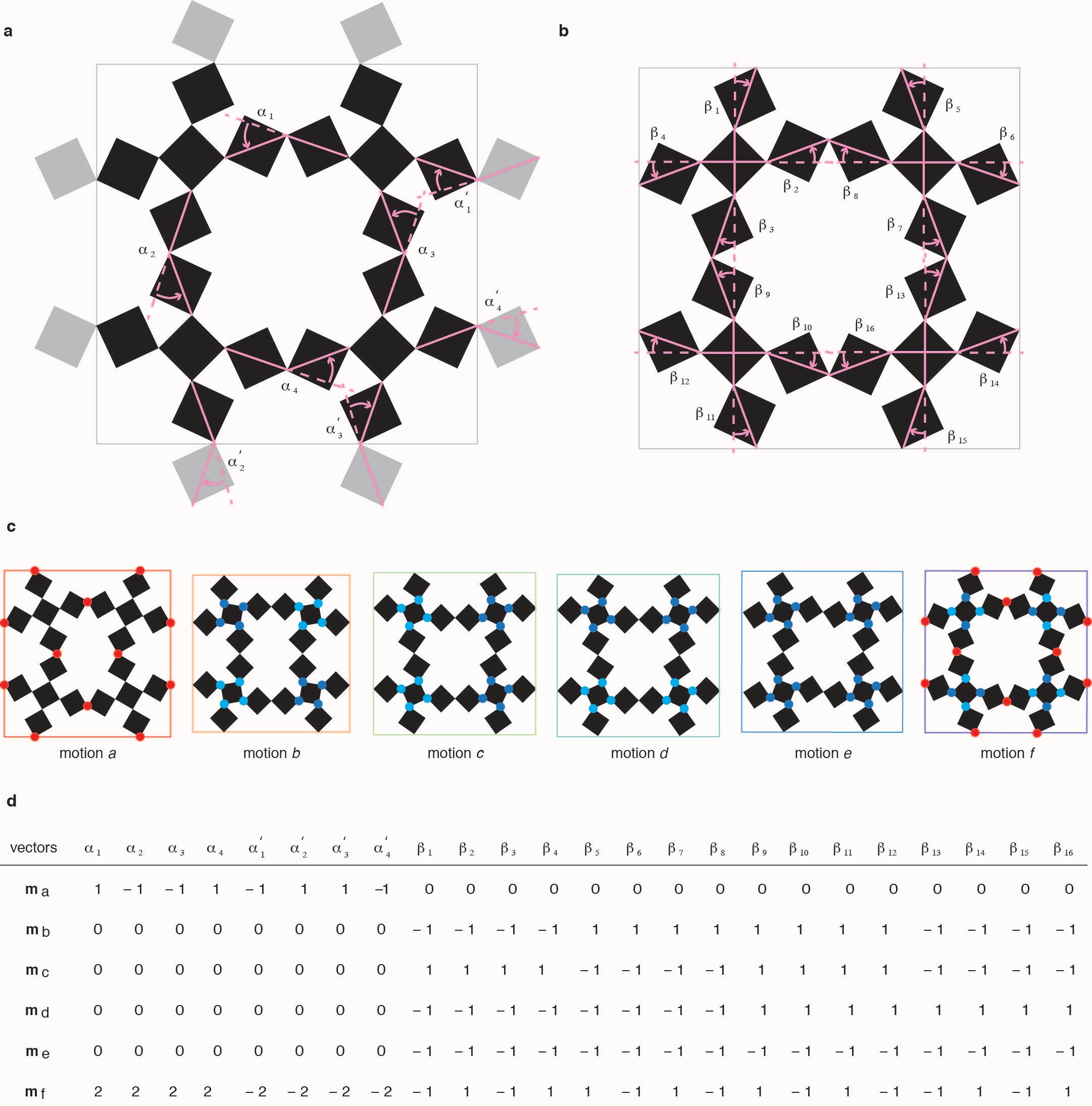}
\end{center}
\caption{{\bf Kinematics and orthogonal base of zero energy motions of a $2\times2$ super cell with periodic square boundary conditions.} (a) Definition of the bending angles $\{\alpha_1,\alpha_2,\alpha_3,\alpha_{4},\alpha'_1,\alpha'_2,\alpha'_3,\alpha'_{4}\}$. The gray squares depict the periodic boundary conditions. (b) Definition of the angles $\{\beta_1,\beta_2,\cdots,\beta_{16}\}$. (c-d) The motions $a-f$ form an orthogonal basis of deformations. (c) Schematic representation. Red colours denote hinging of the $\alpha$-links. Light (dark) blue colours denote clockwise (counter-clockwise) hinging of the $\beta$-links. (d) Vectorial representation. The lines of the table correspond to the vectors $\mathbf{m}_{a,b,\cdots, f}=\{ \alpha_1,\dots,\alpha_4,\alpha'_1,\dots,\alpha'_4,\beta_1,\dots,\beta_{16}\}$ that make up the base of all possible motions $a-f$.}
\label{M10_3}
\end{figure*}

For an $m=2$, $n=2$ super-cell with square periodic boundary conditions we thus find six zero modes. {We define a vector $\mathbf{b}$ describing the bending of the links (See Extended Data Figure~\ref{M10_3}(a-b)) and construct an appropriate orthogonal basis $\mathbf{m}_{a,\dots,f}$}, which we refer to as motions $a-f$ (See Extended Data Figure~\ref{M10_3}(c-d)). We choose motion $a$ and $b$ to coincide with the first step of pathways I (Fig.~1a) and II (Fig.~1b), and use the motions to characterize the soft modes and buckling modes.

\begin{figure*}[h!]
\begin{center}
\includegraphics[width=0.7\linewidth,clip,trim=0cm 0cm 0cm 0cm]{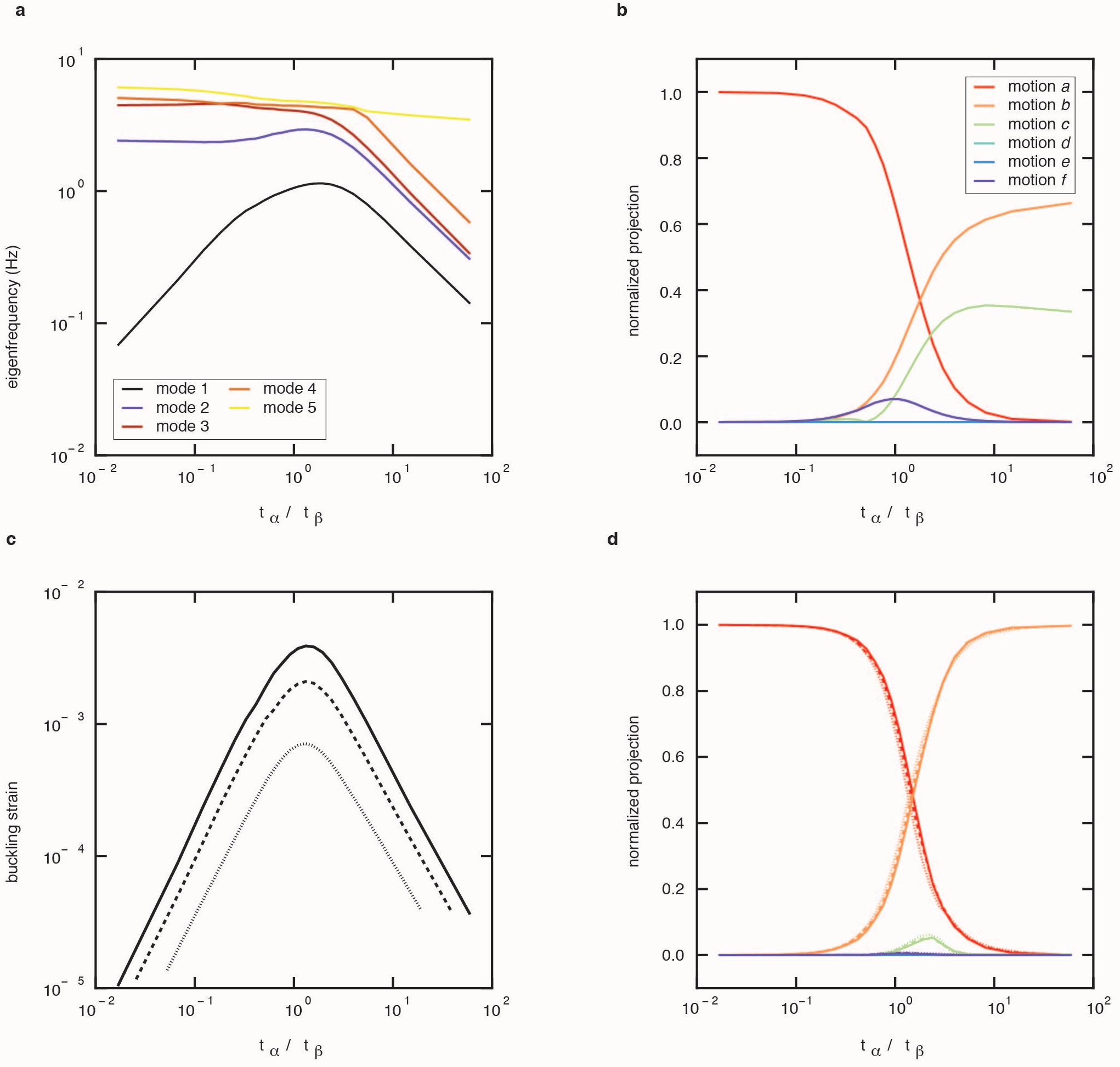}
\end{center}
\caption{{\bf Mode analysis for a $2\times2$ super cell with periodic square boundary conditions.} Unless noted otherwise all data is for $t_{\alpha}\!+\!t_{\beta} =3$ mm, but data for other normalizations look extremely similar.
(a) Eigenfrequencies of the first linear eigenmodes as function of $t_\alpha/t_\beta$.
(b) Normalized projection of the lowest eigenmode onto motions $a-f$. To calculate them, we compute the inner products between the bending angle vectors $\textbf{b}$ and the vectors $\textbf{m}_{a-f}$ defined in Extended Data Figure~\ref{M10_3} and normalize them by the sum of all inner products.
(c) Critical buckling strain as function of $t_\alpha/t_\beta$, for $t_{\alpha}\!+\!t_{\beta} =1, 2$ and $3$ mm  (dotted, dashed and full respectively) --- the value of the critical strain obviously depends on $t_{\alpha}\!+\!t_{\beta}$, but the crossover does not.
(d) Normalized projection of the buckling mode onto motions $a-f$, for $t_{\alpha}\!+\!t_{\beta} =1, 2$ and $3$ mm (dotted, dashed and full respectively). We calculate the normalized projections as in panel (c).}
\label{M7b_5}
\end{figure*}

{\em Linear Modes:} We have calculated the softest eigenmodes at zero strain as function of $t_{\alpha}$
and $t_{\beta}$ using a finite element method (See section ''Numerical simulations" below). We fix $t_{\alpha}\!+\!t_{\beta}$ at 1, 2 and 3 mm and focus on the role of $t_{\alpha}/t_{\beta}$, as we find that the sum $t_{\alpha}\!+\!t_{\beta}$ plays a very minor role for the mode structures. We find that for small values of $t_{\alpha}/ t_{\beta}$, one mode is significantly softer than all others, while for larger values of  $t_{\alpha}/ t_{\beta}$, several modes are in close competition (Extended Data Figure~\ref{M7b_5}(a)). 
{Consistent with our counting argument,} this can be understood by noting that when $t_{\alpha} \ll t_{\beta}$, only motion $a$ is soft, whereas when $ t_{\beta} \ll t_{\alpha}$, motions $b-e$ are all soft, with the eigenmodes mixing these motions.

By characterizing the spatial structure of the softest mode by projecting it onto the motions $a-f$, we obtain the following scenario. First, there is a broad crossover regime where multiple motions are important, with the crossover centered around $t_{\alpha}/ t_{\beta}\simeq 1.5$, consistent with the fact that there are
more $\beta$-links than $\alpha$-links. Second, for small values of $t_{\alpha}/ t_{\beta}$, motion $a$ dominates, while at large values of $t_{\alpha}/ t_{\beta}$, motion $b$ and $c$ both play a role (Extended Data Figure~\ref{M7b_5}(b)). Together, this analysis suggests that for {sufficiently small $t_\alpha/t_\beta$}, where the lowest mode is well isolated and closely corresponds to motion $a$, pathway I can be robustly obtained, whereas the situation for large $t_\alpha/t_\beta$ is more complex, with multiple modes and motions in competition.

{\em Buckling Modes:} To fully understand the role of $t_\alpha/t_\beta$, we have performed a {stepwise} nonlinear analysis and calculated the critical strain at which the first mode becomes unstable, and similarly project the corresponding nonlinear buckling mode onto the six motions $a-f$ (Extended Data Figure~\ref{M7b_5}(c-d)). This data confirms the picture of a crossover at $t_{\alpha}/ t_{\beta}\simeq 1.5$. We note that the buckling modes can almost completely be expressed as a combination of motion $a$ and $b$, with other motions suppressed and essentially irrelevant.

\subsection{Symmetry broken links}
\begin{figure*}[h!]
\begin{center}
\includegraphics[width=0.6\linewidth,clip,trim=0cm 0cm 0cm 0cm]{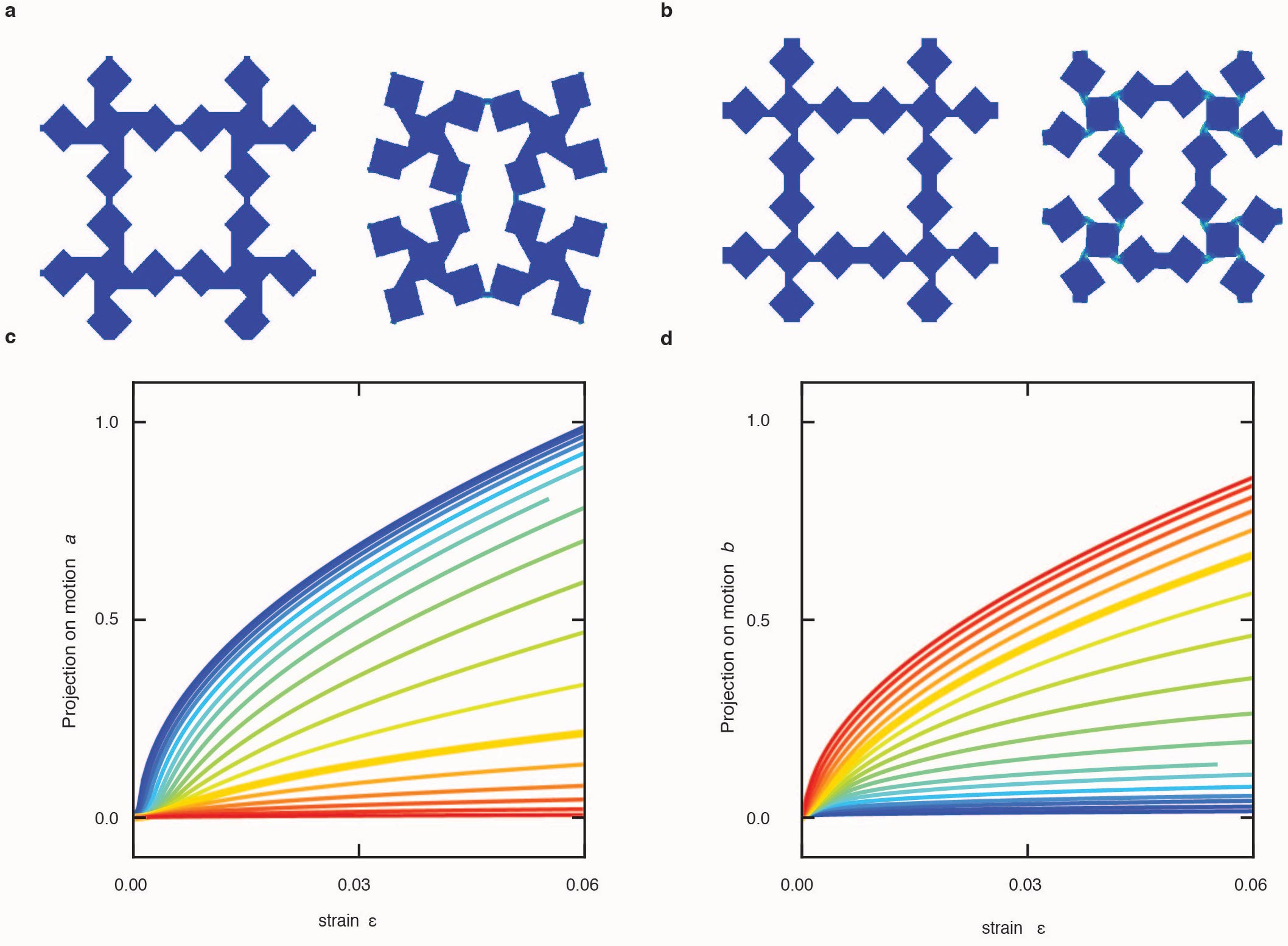}
\end{center}
\caption{{\bf Nonlinear analysis with symmetry broken links on a $2\times2$ super cell with periodic square boundary conditions.} (a-b) Snapshots of the super cell in the undeformed (left) and deformed{---strain $6.6\%$---} (right) state for $t_\alpha=0.9$, $t_\beta=2.1$ (a) and $t_\alpha=2.1$, $t_\beta=0.9$. In both case, the offset of the $\beta$-links is $0.5t_\beta$.
(c-d) Projection of the deformed states $\textbf{b}\cdot \textbf{m}_a$ (c) and $\textbf{b}\cdot \textbf{m}_b$ (d) vs. strain $\varepsilon$ for values of $t_\alpha/t_\beta$ ranging from $0.2$ (blue) to $15$ (red) for $t_\alpha+t_\beta=3$ mm. The case $t_\alpha/t_\beta=2.3$, which is close to the experimental value ($t_\alpha/t_\beta=2.5$, Fig. 2e-h) is highlighted with a thick yellow line.}
\label{M8_7}
\end{figure*}

Execution of pathway II is more difficult than pathway I {presumably because of} the large number of competing modes and motions with deformations of the $\beta$-links. %, and the relatively large ratio of $t_\alpha/t_\beta$ at crossover.
{However, it is possible to select pathway II, even for moderate values of the thickness ratio $t_\alpha/t_\beta $, by introducing a symmetry breaking lateral offset in the $\beta$-links.
We fix the offset magnitude of this additional design parameter at
$|\delta_\beta|=0.5 t_{\beta}$ (so the symmetry breaking is either on or off) and focus on offset patterns consistent with motion $b$. We then run full nonlinear simulations on a $2\times 2$ super cell with square periodic boundary conditions for a range of  values of the ratio $t_\alpha/t_\beta$
while fixing $t_\alpha+t_\beta=3$ mm}.
To illustrate the general scenario, we show snapshots of the undeformed and deformed system for
$t_\alpha/t_\beta = 0.4$ and $t_\alpha/t_\beta = 2.3$ in Extended Data Figure~\ref{M8_7} (a-b). In both cases there is no instability and the deformation is smooth. At large strains we observe the deformed state corresponding to motion $a$ for  $t_\alpha/t_\beta = 0.4$, illustrating that motion $a$ is favoured for small $t_\alpha/t_\beta$ {even in the presence of offsets that promote motion $b$}. However, for larger { $t_\alpha/t_\beta$} the large strain deformation corresponds to motion $b$, illustrating that a combination of symmetry breaking offsets and large $t_\alpha/t_\beta$ allows to initiate pathway II  (Extended Data Figure~\ref{M8_7}(c-d)).

\subsection{Design guidelines and experimental validation}

We now turn our numerical observations on the crossovers between different behaviours as function
of $t_\alpha/t_\beta$, with and without offsets, into practical design guidelines, noting that there are no sharp boundaries. {First,} we expect that pathway I should be observed without offsets whenever $t_\alpha/t_\beta \lesssim 1$, with increasing fidelity for smaller ratio's - for example, for $t_\alpha/t_\beta=0.5$, the overlap between the buckling mode and motion $a$ is more than 90\% (Extended Data Figure~\ref{M7b_5}(d)).
{Second, to obtain pathway II, we suggest to combine an offset and large values of $t_\alpha/t_\beta$.
For the specific offset magnitude used in this paper, $|\delta_\beta|=0.5 t_{\beta}$, and an offset pattern
of alternating chirality, \emph{i.e.} concomitant with motion $b$, our data shows that for {$t_\alpha/t_{\beta} \gtrsim 2$}, motion $b$ dominates; for example, for  $t_\alpha/t_{\beta} =2.3 $, the overlap between the 
nonlinear mode and motion $b$ is already 75\% (Extended Data Figure~\ref{M8_7}(c-d)).}
Finally we note that, experimentally, link thicknesses significantly less than 1 mm are not feasible, and link thicknesses larger than a few mm become comparable with the size of the square elements, thus providing practical constraints on our design space.

{\em Experimental Validation Pathway I:}
We have performed experiments without offset, and
with $t_\alpha/t_\beta= 0.5$ (shown in Fig.~2a-d), {and} $t_\alpha/t_\beta= 0.4$ (not shown) that both show a clear Pathway I, { as well as with $t_\alpha/t_\beta= 1.4$ that shows a disordered Pathway (not shown).} Moreover, earlier experiments with less accurately manufactured 3D printed samples also showed clear Pathway I deformations---Pathway I is robust { for $t_\alpha/t_\beta\lesssim 1$}. In addition, we have performed experiments with an offset pattern {of constant chirality, \emph{i.e.}} consistent with {motion} $e$, for $t_\alpha/t_\beta= 1, 1.2, 1.4, 1.5$ and $2.5$; the former two lead to smoothed versions of pathway I, the latter three to disordered pathways (Fig.~2).

\begin{figure*}[h!]
\begin{center}
\includegraphics[width=0.6\linewidth,clip,trim=0cm 0cm 0cm 0cm]{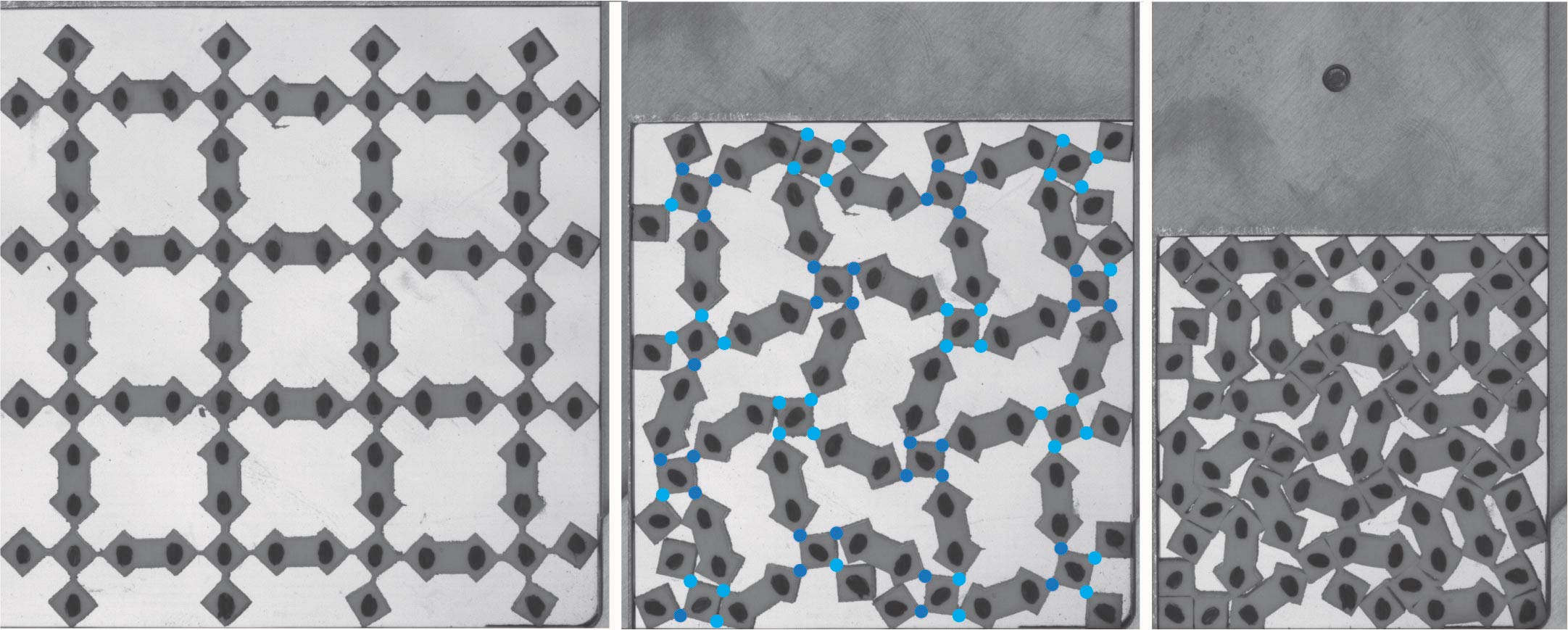}
\end{center}
\caption{{\bf Metamaterial under compression with $t_\alpha=4$ mm and $t_\beta=1$ mm.} Stills of the compression experiment at strains $\varepsilon=0$ (left), $0.24$ (middle) and $0.52$ (right). On the middle panel, the light (dark) blue dots indicate clockwise (counter-clockwise) hinging of the $\beta$-links, similar to motion $b$ that initiates pathway II, yet with significant disorder near the boundaries that penetrate into the bulk.}
\label{M11_6}
\end{figure*}

{\em Experimental Validation Pathway II:}
We have attempted to experimentally obtain pathway II %with two samples
without offset and $t_\alpha/t_\beta= 4$. Despite the fact that a large part of the sample follows motion $b$ these samples also display a significant amount of disorder, seemingly nucleating near the boundaries, that prevent a successful execution of pathway II (Extended Data Figure~\ref{M11_6}). We therefore have performed experiments with an offset  {$|\delta_\beta|= 0.5 t_{\beta}$} with an alternating chirality consistent with {motion} $b$ (Fig.~2). We find that pathway II can be obtained for a moderate value of $t_\alpha/t_\beta = 2$
(not shown) and  $t_\alpha/t_\beta = 2.5$ (Fig.~2).

\subsection{Numerical simulations}
For the finite elements simulations of a representative metamaterial element we use the commercial software Abaqus/Standard.

\emph{Model definition}. We model $2\times 2$ super cells of our metamaterials varying the parameters $t_\alpha$, $t_\beta$ and the offset $\delta_\beta$ (Extended Data Figure~\ref{M4_8}) using a neo-Hookean energy density as a material model, using a shear modulus, G = $2.7$ MPa and bulk modulus, K = $133$ GPa (or equivalently a Young's modulus E = $8.0$ MPa and Poisson's ratio $\nu=0.49999$) in plane stress conditions with hybrid quadratic triangular elements (Abaqus type CPS6). We construct the mesh such that the thinnest parts of the samples with at least two elements across. As a result, the metamaterial unit cells have approximately $10^4$ triangular elements.

\emph{Boundary conditions}. To implement periodic square boundary conditions, we define constraints on the displacements of all the nodes at the horizontal and vertical boundaries of the unit cell~\cite{Coulais_IJSS_2016}. In addition, we ensure that the periodic boundary condition remain square-shaped.
%\mvh{REMOVE THIS: To this end: (i) we impose that the horizontal and vertical boundaries undergo the same displacement; (ii) we add two additional constraints that prevent shear deformations by imposing that the boundaries remain straight. These two constraints, which are defined on the central node of the two $\alpha$-links at the upper horizontal (right vertical) boundary, prevent any relative vertical (horizontal) displacement between the centre of the $\alpha$-links, yet allow rotations of these beams.}

\noindent We perform three types of analysis:

\emph{Linear eigenmodes analysis:}
We calculate the lowest eigenmodes and their eigenfrequencies (Extended Data Figure~\ref{M7b_5} (a-b)).

\paragraph{Nonlinear bifurcation analysis:} We perform a stepwise nonlinear stability analysis to determine the bifurcation point with a relative accuracy of $5\times 10^{-2}$ (Extended Data Figure~\ref{M7b_5} (c-d)).

\paragraph{Nonlinear compression using imperfection:} To study metamaterials whose $\beta$-links have an offset $\delta_\beta$, we compress the structure up to a strain $6.6\%$ (Extended Data Figure~\ref{M8_7}).

\subsection{Experimental techniques}

We used 10 mm-thick cast sheets of silicone rubber (shore 80A, Silex Silicones LTD, Young's modulus $E=8$ MPa), that is sufficiently stiff to ensure  minimal affection by gravity, and exhibits minimal viscous and creeping effects. We fabricated the samples using a waterjet cutter, gluing the rubber to 10 mm thick plywood boards to ensure precise cutting. We then marked each internal square (side $4.5$ mm) of the sample with an ellipse ($2.5$ mm$\times4$ mm) for detection purposes.

\begin{figure*}[!t]
\begin{center}
\includegraphics[width=0.6\linewidth,clip,trim=0cm 0cm 0cm 0cm]{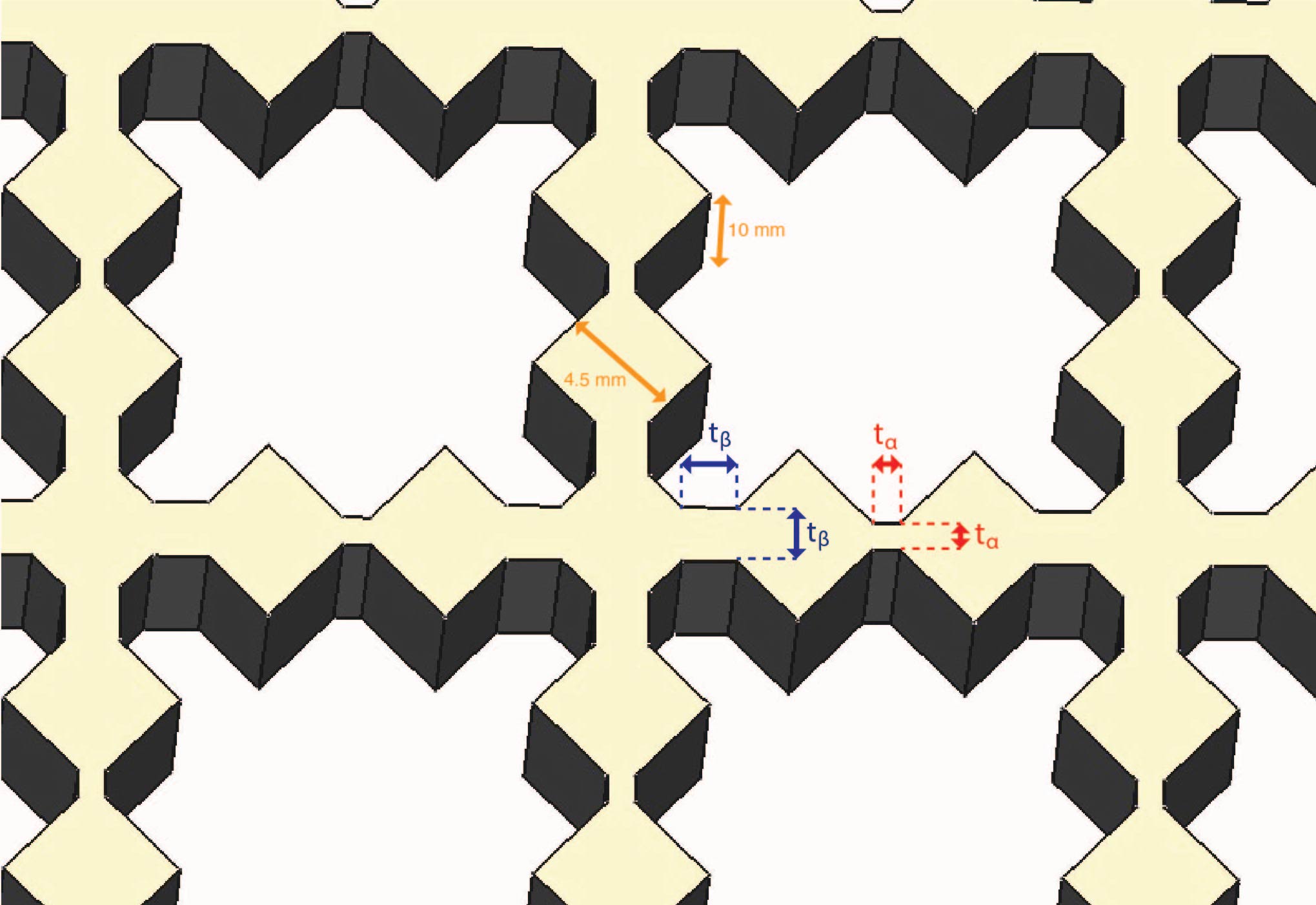}
\end{center}
\caption{{\bf Sample geometry.} Zoom-in of a computer assisted design of the rank II metamaterial (here for the sample shown in Figs.2 (a-d).}
\label{M4_8}
\end{figure*}

The hinges between each square consist of rectangular cuboid beams of height $10$~mm (same at the sheet thickness) and of lateral thickness
%and length NOT CORRECT FOR OFSET
 $t_\alpha$, $t_\beta$ or $t_\gamma$. We note that, irrespective of the beams, the (virtual) corners of the squares precisely meet in a point --- the beams add material to this idealized design (Fig.\ref{M4_8}).
The thicknesses of $t_\alpha$, $t_\beta$, $t_\gamma$ range from $0.8$ mm to $4.0$ mm, depending on the nature of the link ($\alpha$, $\beta$ or $\gamma$) and on the sample (see main text and Methods for specifications). In some runs, we laterally offset the beams by $\delta_\beta = t_{\beta}/2$, leading to triangular shaped links. {The rank II (rank III) metamaterials shown in Figs. 1,2 (Fig. 3) consist of $4\times4$ ($3\times3$) unit cells, namely $80$ ($225$) internal squares.}

\begin{figure*}[!h]
\begin{center}
\includegraphics[width=0.6\linewidth,clip,trim=0cm 0cm 0cm 0cm]{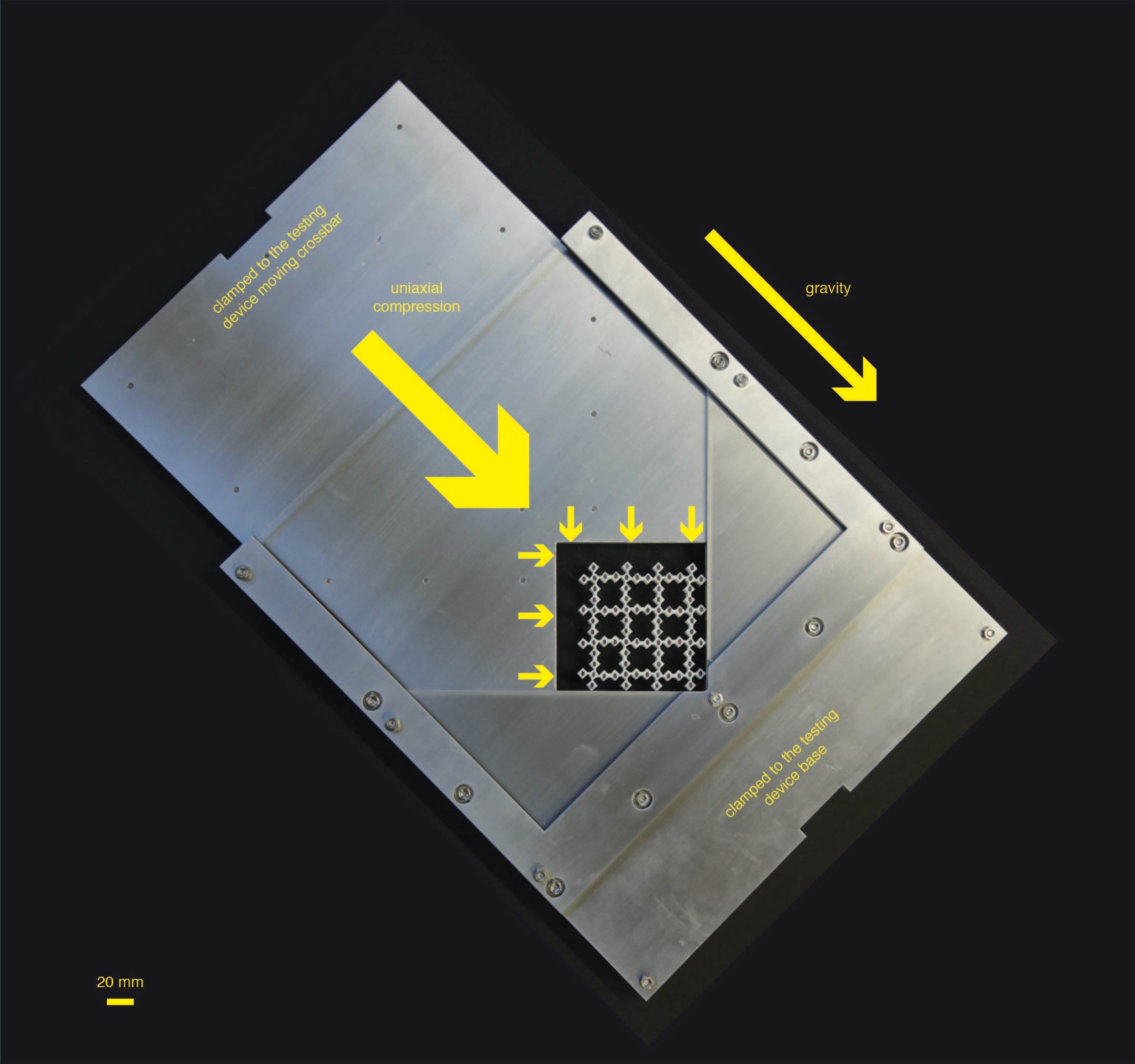}
\end{center}
\caption{{\bf Compression device.}
Custom made compression fixture to apply biaxial compression using a uniaxial testing device. }
\label{M5_9}
\end{figure*}

The samples were tested under equi-biaxial compression. To this end, we used a custom made aluminium V-shaped press (See Extended Data Fig.~\ref{M5_9}) that was carefully positioned and aligned in a universal testing machine (Instron 3366) equipped with a $1000$ N load cell, thus allowing to impose a compressive displacement within a $10~\mu$m accuracy and to record the force within a $0.1$ N accuracy. We used fine powder to reduce the friction at the boundaries as much as possible.

The pictures were recorded using a high-resolution CMOS camera ($3858$ px$\times 2764$ px, Basler acA3800-14um) equipped with a 75 mm (50 mm) prime lens (KOWA, LM50HC 1'' and LM75HC 1'', respectively) for the rank two (rank three) metamaterials paired with custom made LED-based front and back light systems. In order to minimise reflections from the front light, the V-press glass windows are coated against reflections.

\subsection{Image tessellation techniques}

The images were processed through standard image tessellation and tracking techniques to extract the positions of the squares as well as the bending angles of the links connecting the squares. In particular, through a semi-automatic custom-made tracking algorithm, special attention was devoted to obtain space-time trajectories of nearly all squares and bending angles (See Supplementary Videos).
Inevitable detection errors occurred close to the edges of the compression cell or when squares come in contact and led to flickering of the square positions and orientation and therefore of the measurement of the bending angles. To tackle this issue, events where the angles fluctuates more than $10^\circ$ between two consecutive frames have been filtered out. We verified that the value of the filtering threshold has a minor effect on the data only.

\begin{figure*}[!b]
\begin{center}
\includegraphics[width=0.6\linewidth,clip,trim=0cm 0cm 0cm 0cm]{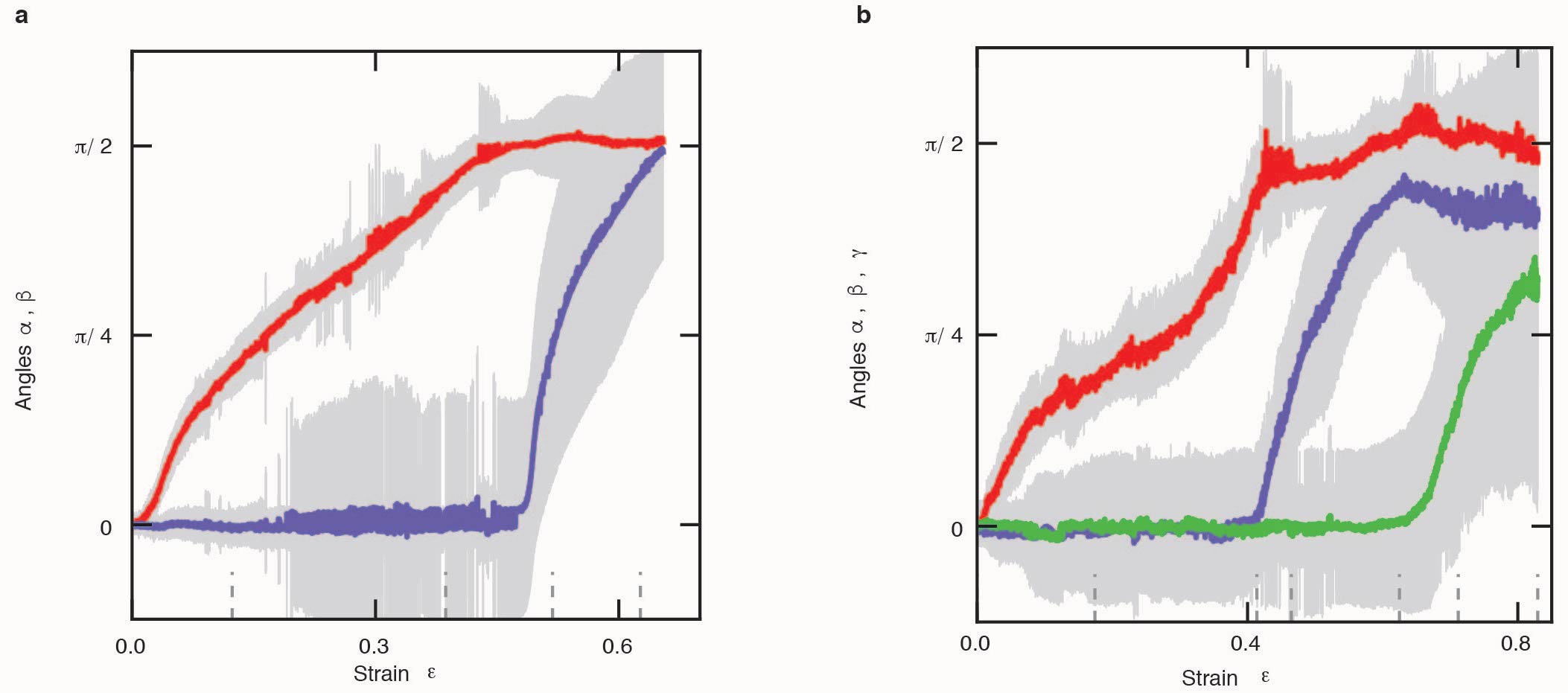}
\end{center}
\caption{{\bf Effect of boundaries on angles.}
Bending of the links for larger regions of the sample than shown in the main text. (a)
 $\alpha$ (red), $\beta$ (blue) vs. strain $\varepsilon$ for the central 64 squares (excluding the outermost 16 squares); grey indicates one standard deviation and the angles are based on the  same run as in Fig.~2a-d of the main text.
 (b)  $\alpha$ (red),  $\beta$ (blue) and $\gamma$ (green) vs. strain $\varepsilon$ for the central 165 squares (excluding the outermost 60 squares); grey indicates one standard deviation and the angles are based on the same run as in Fig.~3 of the main text.}
\label{M6_10}
\end{figure*}

In addition, as discussed in the main text, friction and frustration at the boundary inevitably cause distortions and misfoldings, whose description goes beyond the scope of this work. Therefore, in order to compute the bending angles shown in Figs. 2d,h,l,p (Fig. 3), we restrict our attention to a region of interest comprising the 28 (29) most central squares for the rank II (rank three) metamaterial. In Fig.~\ref{M6_10} we show the same data, now calculated over a much wider area, comprising 64 (respectively 165 squares)---even though the scatter is larger, the main trends and multi-step nature are also apparent in this data.

\subsection{Sequential pathways with alternative topologies}

\begin{figure*}[h!]
\begin{center}
\includegraphics[width=\linewidth,clip,trim=0cm 0cm 0cm 0cm]{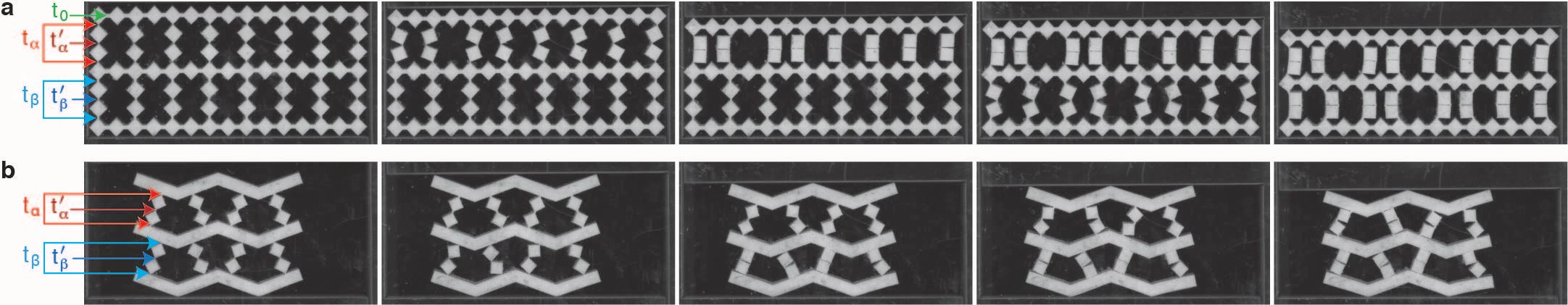}
\end{center}
\caption{{\bf Alternative Topologies.} (a) Linked squares with edges 4.5 mm, and link thicknesses
$t_0=1.35$ mm, $t_\alpha=0.45$ mm, $t_{\alpha'}=0.45$ mm,
$t_\beta=0.9$, $t'_\beta=0.68$ mm showing a two-step folding pathway under uniaxial compression. (b) Structure of linked squares with edge 4.5 mm, and link thicknesses
$t_\alpha=0.45$ mm, $t_{\alpha'}=0.45$ mm, $t_\beta=0.9$ mm, and $t'_\beta=0.9$
 mm connected by slanted bars of thickness $4.5$ mm showing a two-step folding pathway under uniaxial compression.}
\label{M9_11}
\end{figure*}

To demonstrate that sequential pathways are not limited to the hierarchical structures under biaxial compression shown in the main text, we have constructed two alternative geometries that show a two-step pathway under uniaxial compression, using the same rubber and water-jetting technique as for the samples shown in the main text (see Extended Data Figure~\ref{M9_11}). The basic idea is to couple two groups of links with different buckling thresholds in series, so that under compression these two types of links buckle in sequence. In one geometry, we use a diluted square lattice consisting of two rows of coupled columns with different link thicknesses (Extended Data Figure~\ref{M9_11}(a)). Under compression, we observe that the columns with the thinnest links buckle first, then fold up until self-contacts are created, and then trigger the buckling of the other columns with thicker links. We note that as this structure is soft to lateral shear, we use lateral sliding boundaries. In the second geometry, we use a variation that removes the soft shear modes, so that no lateral boundaries are required; also here, a clear two-step sequence is observed (Extended Data Figure~\ref{M9_11}(b)). These alternative designs demonstrate that the combination of buckling and self contacts can generate sequential pathways in a variety of structures.

\textbf{Data and code availability.} The data and codes that support the plots within this paper and other
findings of this study are available from the corresponding author upon request.

%\clearpage
\section{Extended Data}

\end{document}